\documentclass[preprint,aps]{revtex4}

\input epsf
\usepackage{dcolumn}

\begin{document}

\title{A Theory of Fluctuations in Stock Prices}

\author {\'Angel L. Alejandro-Qui\~nones$^{\dagger}$,~Kevin E. Bassler$^{\dagger}$,~Michael Field$^\ddagger$,~Joseph L. McCauley$^{\dagger}$,~
Matthew Nicol$^\ddagger$,~Ilya Timofeyev$^\ddagger$,~Andrew T\"{o}r\"{o}k$^\ddagger$,~and Gemunu H. Gunaratne$^{\dagger}$}

 \altaffiliation[Also at ] {The Institute of Fundamental Studies,
                Kandy 20000, Sri Lanka}
 \email {gemunu@uh.edu}

 \affiliation{~$^{\dagger}$Department of Physics,
                University of Houston,
                Houston, TX 77204}

 \affiliation{~$^{\ddagger}$Department of Mathematics,
                University of Houston,
                Houston, TX 77204}


\begin{abstract}
The distribution of price returns for a class of
uncorrelated diffusive dynamics is considered. The
basic assumptions are (1) that there is a ``consensus"
value associated with a stock, and (2) that the rate
of diffusion depends on the deviation of the stock price
from the consensus value.  We find an analytical expression
for the distribution of returns in terms of the diffusion
rate, when the consensus value is assumed to be fixed in
time.  The analytical solution is shown to match computed
histograms in two simple cases.  Differences that result
when the consensus value is allowed to change with time
are presented qualitative explanations.

\end{abstract}

\maketitle

\section{Introduction}

Statistical analysis of a wide range of physical and other processes
have shown the presence of non-Gaussian distributions.  They include
temperature fluctuations in hard turbulence~\cite{castaing}, diffusion
in inhomogeneous media~\cite{havAavr,hanAkeh}, and price variations in
financial markets~\cite{manAsta,dacAgen}. One common
characteristic in these distributions is the presence of exponential
or power-law tails, signifying a more frequent occurrence of large deviations
than expected from a collection of independent, identically distributed
events~\cite{feller}. The width of the distributions have also been shown
to scale as the time interval during which the fluctuations
occur.  Based on these properties, it has been proposed that Levy distributions
be used to describe fluctuations in the underlying processes~\cite{geoAbou,schesinger}.
In this paper, we propose an alternative explanation for the non-Gaussian
distributions, namely a non-uniform diffusion rate~\cite{vankampen,friApei}.
The discussion here is based on fluctuations in financial markets.

Consider a stock whose price at time $t$ is given by $S(t)$.
Most financial market analyses are conducted in terms of the ``return"
of stock, $x(t)=\ln[S(t)/S_0]$, where $S_0$ is a reference price.
Studies have shown that successive fluctuations in return are
uncorrelated~\cite{dacAgen,arnAbou}.
Most theoretical analysis assume further that successive events are
statistically independent~\cite{feller}. The traditional theoretical model uses
a simple discrete random walk in the logarithm of price to describe
the dynamics of an asset in a market. This model results in a normal,
or Gaussian distribution of price returns~\cite{osb}. 
On the other hand, empirical data clearly show that the 
price return distribution of real markets deviates significantly from
a Gaussian, especially far from the mean~\cite{mand}.  
(For recent reviews see \cite{manAsta,dacAgen,bouApot,gop,ple,leeAlee,coronel}.)
In particular, 
some detailed studies~\cite{gop,dacAgen,ple} have found that the tails of
the distribution have an asymptotic power law decay, typically with exponents
larger than 2, while others~\cite{macAgun, bouApot} have found that the tails 
of the distribution are better described by an exponential decay.  

It has been conjectured that the non-normality observed in real financial
markets can be explained by assuming that although successive events may be
uncorrelated, the rate of trading depends on the price of the stock~\cite{macAgun}.  
In particular, it is assumed that there is a ``consensus"
value $\bar{x}(t)$ of the return, at which point the stochastic fluctuations
are minimum, and the level of the fluctuations increase as the return
moves away from $\bar{x}(t)$.  In other words, the rate of transactions
depend on the return. In particular, a large change in price is likely
to be followed by large fluctuations.  

Here, we explore the implications of the conjecture by studying 
variations of the traditional model. Like the traditional model,
we use an uncorrelated random walk to represent the dynamics of
price returns. However, unlike the former, where the rate of trading
is independent of price, our model assumes a rate which increases
the deviation of the stock price from $\bar{S}(t)$. It will be
demonstrated that this modification in the dynamics can reproduce
the range of non-Gaussian behavior of the price return distributions
observed empirically in real markets.

If an uncorrelated random walk models fluctuations of price returns,
then the continuous dynamics is a diffusive process. In models considered here,
the diffusion coefficient $D(x,t)$ is assumed to depend on the time
elapsed $t$ and the return $x(t)$.  It is assumed further that the
dependence on $x(t)$ is a function of its deviation from the consensus
return $\bar{x}(t)=\ln[\bar{S}(t)/S_0]$.

The remainder of the paper is structured as follows.
First, exact analytical results are presented for 
the distribution of price returns, valid for an arbitrary
functional form of the diffusion coefficient, when $\bar{S}$
is fixed in time. The analytical solutions for the distributions
are obtained by solving the Fokker-Planck equation for the
corresponding diffusive process. Then, two specific functional forms
for $D(x,t)$ are considered. The first form is a
piecewise linear function of $u = x / \sqrt{t}$.
While the second is a quadratic function of $u$.
The form of the distributions are obtained analytically.
Results from simulations of the corresponding discrete random walks
are also presented and shown to be consistent with the analytical
forms. Subsequently, we provide results from computations
for a model with time dependent consensus value.
In conclusion, we summarize the results and discuss their importance 
for understanding the behavior
of real markets.

\section{Exact analytical solution for the return distribution}

To obtain an analytical expression for the price return distribution 
$W(x,t)$ of the diffusive processes we are considering, note that 
it satisfies the Fokker-Planck equation~\cite{chan,fokApla}
\begin{equation}
\frac{\partial W}{\partial t} = -R(t) \frac{\partial W}{\partial x} + \frac{1}{2
}
  \frac {\partial^2} {\partial x^2} (DW),
\label{F-P}
\end{equation}
where $D\equiv D(x,t)$ is the diffusion coefficient~\cite{levi}, and
$R(t)$ is a (time-dependent) drift rate~\cite{other}.  
For simplicity, we assume $R(t) = 0$ for the rest of this analysis.  
However, the case of non-zero $R(t)$ can also be treated using a simple
coordinate transformation $x' = x - \int_{0}^{t} R(t')~dt'$.

A normalizable solution to Eq.~(\ref{F-P}), consistent with empirical
investigations of financial markets~\cite{manAsta,dacAgen,conApot},
can be found by assuming that the distribution of returns has
the scaling form
\begin{equation}
W(x,t) = \frac{1}{t^{\eta}} F(u).
\label{scaling}
\end{equation}
Here $u=x/t^{\eta}$, and $\eta$ is the self-similarity exponent~\cite{manAsta}.
We also assume that the diffusion rate is a function of $u$.
This scaling hypothesis leads a unique value for $\eta$,
which can be seen by noting that, using it,
Eq.~(\ref{F-P}) becomes
\begin{equation}
  -\frac{\eta}{t^{\eta+1}} F(u) - \frac{\eta}{t^{\eta+1}} u F'(u)
      = \frac{1}{2} \frac{1}{t^{3\eta}} (DF)^{\prime\prime}(u).
\label{wander}
\end{equation}
Consequently $\eta=1/2$, a value which consistent with conclusions from
empirical studies of real markets. Then Eq. (\ref{wander}) simplifies to
\begin{equation}
	[D(u) F(u)]'' + [u F(u)]' = 0,
\label{F-Psimp}
\end{equation}
which can be integrated to
\begin{equation}
	[D(u) F(u)]' + u F(u) = Const.
\label{F-Psimp2}
\end{equation}
If $D(u)$ is symmetric about $u = 0$ and the diffusion process
starts at the origin, then $F(u)$ will also be symmetric about $u = 0$.
Under these conditions, both terms in the LHS of Eq. (\ref{F-Psimp2})
are anti-symmetric about $u=0$, and thus $Const = 0$.  
Therefore, 
\begin{equation}
	D(u) F(u)' = -[u + D(u)'] F(u)
\label{F-Psimp3}
\end{equation}
which has a general solution of the form
\begin{equation}
	F(u) = \frac{1}{D(u)}~exp \left[ -\int^{u}\frac{\bar u}{D(\bar u)}\,d \bar u \right].
\label{F-Psimp4}
\end{equation}

As an example, consider a constant diffusion coefficient
$D(x,t)=D_0$.  In this case, Eq. (\ref{F-Psimp4}) gives a
solution of the form
\begin{equation}
	F(u) = C_0~exp \left[ -\frac{1}{2D_0}u^2 \right],
\label{gaussian}
\end{equation}
which is the well-known result for the 
traditional model of distribution of returns.

\section{Static consensus price $\bar{S}$}

In this section we present two models that use
a non-constant diffusion coefficient $D(u)$
and take the consensus value $\bar{S}(t)$ to be fixed,
without loss of generality, at $\bar{S}(t)=S_0$.
For both cases we present numerical and analytical results.

\subsection{Piecewise Diffusion}

The first form of $D(u)$ we consider is a piecewise linear function
\begin{equation}
   D(u)=D_0(1 + \epsilon |u|),
\label{lindiff-coef}
\end{equation}
where $D_0$ and $\epsilon$ are constant parameters.  It should be noted
that $D_0$ can be eliminated by a suitable rescaling of time.
The exact solution to Eq. (\ref{F-P}),
obtained using Eq.~(\ref{F-Psimp4}),
is
\begin{equation}
F(u)=C_0~exp\left[ -\frac{|u|}{D_0 \epsilon} \right] \left(\epsilon |u| + 1\right)^{\alpha-1}, 
\label{lindiff-dist}
\end{equation}
where $\alpha = 1/(D_0 \epsilon^2)$, the constant $C_0$ which normalizes $W(x,t)$ is given by
\begin{equation}
C_0 = \frac{[1/(D_0 \epsilon e)]^{\alpha}}{2 \sqrt{t}~\Gamma[\alpha,\alpha]},
\label{lindiff-norm}
\end{equation}
and
\begin{equation}
\Gamma [a,z]=\int_{z}^{\infty} p^{a-1} e^{-p}dp
\label{inc-gamma}
\end{equation}
is the incomplete Gamma function.
	  
In the limit that $\epsilon$ vanishes, $W(x,t)$ becomes a Gaussian.
This can be seen from
\begin{equation}
\ln F(u) \sim \left( \frac{1}{D_0 \epsilon^2} - 1\right) \ln (1 + \epsilon |u|) - \frac{|u|}{D_0 \epsilon}
	\sim \frac{-u^2}{2 D_0} + O(\epsilon)
\label{limit}
\end{equation}
and hence 
\begin{equation}
\lim_{\epsilon \to 0} F(u) \sim exp[-u^2 / 2D_0]
\label{limit2}
\end{equation}
This is because, in that limit, 
the diffusion coefficient (\ref{lindiff-coef}) is a constant.
As $\epsilon$ increases the tails of the
distribution decay slower.  We refer the reader to Ref.~\cite{macAgun}
for a more detailed study of the special case $\epsilon = 1/\sqrt{D_0}$
when $F(u)$ is an exponential distribution.
	   
\begin{figure}[htbp]
\protect\vspace*{-0.1cm}
\epsfxsize = 3.0 in
\centerline{\epsfbox{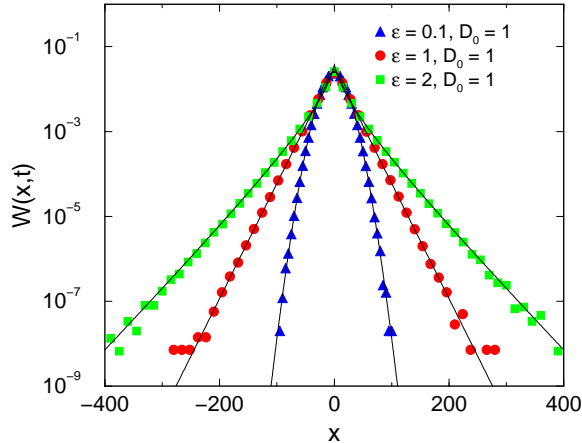}}
\protect\vspace*{0.2cm}
\caption{Distribution of returns for a piecewise linear diffusion coefficient, 
Eq.~(\ref{lindiff-coef}), with $D_0=1$. 
The shape of the distribution changes with the parameter $\epsilon$. 
Notice the special cases of $\epsilon=0$ where the distribution 
is Gaussian, and of $\epsilon=1$ where the distribution is exponential. 
The solid lines represent the analytical solution of Eq.~(\ref{lindiff-dist}),
and the data points are the results of the random walk simulations with
$10^6$ independent walks, each one lasting a time $t=256$.}
\label{figure1}
\end{figure}

We simulated the price returns using random walks with steps
of unit size occurring at non-constant time intervals.
The time between steps is $1/D(x,t)$~\cite{chan}, where
$D(x,t)$ was calculated at every time step.
The simulations consisted of many independent walkers, each of which
started at the origin, and randomly chose the direction of each event to be
either an increase or a decrease with equal probability.  The walks continued
until a maximum time was reached. Fig.~\ref{figure1} compares
the analytical and simulation results, showing good agreement between them.

\subsection{Quadratic Diffusion}

The second form of $D(u)$ we consider is,
\begin{equation}
D(u)=D_0(1 + \epsilon u^2).
\label{quaddiff-coef}
\end{equation}
It this case, 
the solution to Eq. (\ref{F-P})
obtained using Eq.~(\ref{F-Psimp4}),
is
\begin{equation}
   F(u)=\frac{C_0}{(1 + \epsilon u^2)^{1 + \beta}},
\label{quaddiff-dist}
\end{equation}
where $\beta = 1/(2 D_0 \epsilon)$ and the normalization constant for $W(x,t)$ is
\begin{equation}
   C_0 = \frac{\Gamma[1+\beta]}{\sqrt{t/\epsilon} \; 
   \Gamma[\frac{1}{2}] \; \Gamma[\frac{1}{2} + \beta]}.
\label{quaddiff-norm}
\end{equation}

\begin{figure}[htbp]
\protect\vspace*{-0.1cm}
\epsfxsize = 3.0 in
\centerline{\epsfbox{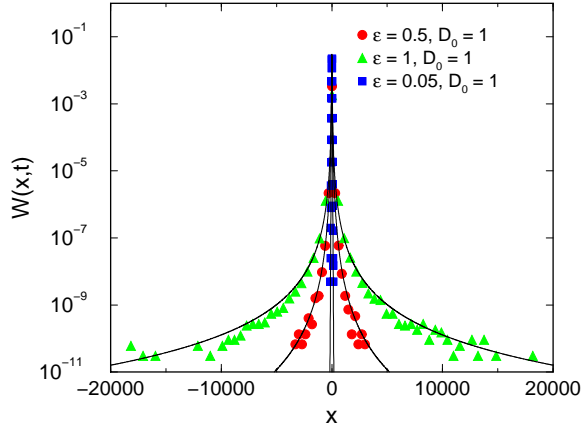}}
\protect\vspace*{0.2cm}
\caption{Distribution of returns using the quadratic diffusion coefficient
in Eq.~(\ref{quaddiff-coef}). Note that the distribution is not exponential,
but instead has fat tails. The solid lines represent the exact solution to
Eq. (\ref{quaddiff-dist}), and
the data points represent the results from the simulation. 
For $\epsilon = 0.05$ $2\times 10^7$ walks were simulated, 
$5\times 10^7$ for $\epsilon = 0.5$, and $6 \times 10^7$ for $\epsilon = 1$.
The final time used in each case was $t=256$.}
\label{figure2}
\end{figure}

This result is plotted in Fig.~\ref{figure2}, where it
is compared to the results of the corresponding discrete random
walk simulation for different values of $\epsilon$, with $D_0=1$. 
The simulations were performed as in the case of
piecewise linear diffusion, except that in this case $D$ is given
by Eq.~(\ref{quaddiff-coef}). Note also that the results from the
simulation are again consistent with the analytical solutions.

\begin{figure}[htbp]
\protect\vspace*{-0.1cm}
\epsfxsize = 3.0 in
\centerline{\epsfbox{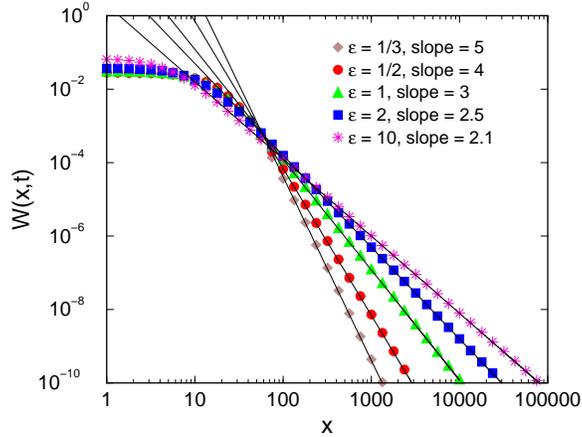}}
\protect\vspace*{0.2cm}
\caption{Log-log plot of the distribution of $x$ using a quadratic $D(x,t)$
showing that the power-law tails of $W(x,t)$ can 
have exponents ranging from
2 upward. The data points represent the analytical solution, from Eq.~\ref{quaddiff-dist}.  
The tail of each case is fitted with a straight line having a slope equal 
to the tail exponent.}
\label{figure3}
\end{figure}

As before, the 
return distribution also becomes a Gaussian in this case 
when $\epsilon$ vanishes.  
However, as $\epsilon$ increases, the tails 
of $W(x,t)$ become power-law distributed.
This behavior can better be appreciated in Fig.~\ref{figure3} where a
log-log plot for different values of $\epsilon$ is presented.
In the limit of $\epsilon \rightarrow \infty$ the tails of the distribution are 
well fitted by a power-law with exponent 2. 
Meanwhile, as 
$\epsilon \rightarrow 0$ 
the tail can also be fitted with 
a power-law, but with an exponent whose value increases and $\epsilon$ 
decreases. However, the fit is good over a range that shrinks as $\epsilon$
decreases. This is expected since the distribution becomes Gaussian 
in the limit 
$\epsilon \rightarrow 0$. 
It is important to point out that as $\epsilon$ is decreased from 
$\infty$ to $0$ the exponent observed in the tail varies from 2 to $\infty$.
Thus, these results reproduce the empirical observations of real markets
that find fat tailed price return distributions with exponents ranging
from 2 upward. Exponents as large as 7.5 have been reported,~\cite{dacAgen},
but at these values the results have large error bars. 
This is because large exponents are found when the time scale is increased,
and the amount of data samples used in the analysis decreases.

Fat tails $f(x,t) \sim |x|^{-\alpha - 1}$ for $|x|>>1$, in the
range $2 < \alpha < \infty$ are also generated by symmetric L\'{e}vy distributions
\cite{plevy},
\begin{equation}
f(x,t)=\frac{1}{2 \pi} \int_{-\infty}^{\infty} dk~e^{ikx-|k|^{\alpha} \Delta t},
\label{levy}
\end{equation}
when $\langle \Delta x^2 \rangle = \Delta t^{2/\alpha}$ for $\alpha > 2$, but this implies
fractional Brownian motion with Hurst exponent $H<\frac{1}{2}$.  Here, there
is no Langevin equation with in $x$ ($D(x,t)$ doesn't exist), and the efficient
market hypothesis is violated~\cite{hull}.

\section{Dynamic consensus price $\bar{S}$}

As has already been mentioned, the price of the stock will fluctuate about the
consensus value $\bar{S}$.  However, contrary to what has thus far been 
considered, 
there is no reason to expect that the consensus value should stay fixed.
In fact, it is plausible that with every trade $\bar{S}(t)$ 
will shift by a small amount toward the 
value of the current price $S(t)$, or  
equivalently, that the consensus value of returns $\bar{x}(t)$ will
shift toward the value of the current return $x(t)$.
Of course, the diffusion constant will change with the consensus value changes
to $D(x-\bar{x},t)$ in order to keep its minimum at $x=\bar{x}$.

In this section, results of
random walk simulations which utilize the non-constant
diffusion coefficients considered in the previous section,
and which allow for the consensus value to change, are presented.
A very simple dynamics for the value the consensus return $\bar{x}$ is considered; 
the change in the value at each time step, $\Delta \bar{x}$, 
is assumed to be proportional to the difference in
the current return value and its current value,
\begin{equation}
\Delta \bar{x}(t) = k[x(t) - \bar{x}(t)],
\label{shift}
\end{equation}
where $k>0$ is a constant that we will assume to be small.
There are two essential differences between the simulations
discussed in section III and those in this section.
First, as mentioned above the diffusion constant used is $D(x-\bar{x},t)$ 
instead of $D(x,t)$.
Therefore, the time between steps becomes $1/D(x-\bar{x},t)$.
Second, the value of $\bar{x}$ is varied dynamically
using Eq.(\ref{shift}). 

The simulations again begin with the consensus price of the stock equal to
its initial value $\bar{S}(0)=S_0$, and therefore the initial 
value of the consensus
return vanishes $\bar{x}(0)=0$. 
Subsequently, $\bar{x}$ will fluctuate around its
initial value. Of course, $x(t)$ will also fluctuate about the origin. When
the value of $x(t)$ is near the origin, it is often the case that 
$|x(t)| \approx |\bar{x}(t)|$.
This causes the peak in the distribution of returns $W(x,t)$ to smear out.

\subsection{Linear Diffusion}

\begin{figure}[htbp]
\protect\vspace*{-0.1cm}
\epsfxsize = 3.0 in
\centerline{\epsfbox{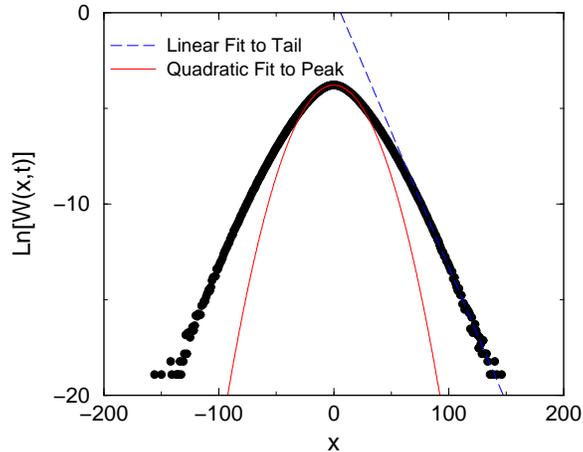}}
\protect\vspace*{0.2cm}
\caption{Distribution of returns with linear diffusion and dynamic consensus value.
The center of the distribution of $x(t)$ is well described by a quadratic function. The curve shown was fit to the 31 points in the
center.  The tails of the distribution have an exponential decay.
A straight line was used to fit the tails.
The figure shows results for $\epsilon = 1$, $D_0 = 1$ and $k = 0.01$. $1.6 \times 10^8$ walks, each with a final time of $t = 256$, were used in the simulation.}
\label{figure4}
\end{figure}

\begin{figure}[htbp]
\protect\vspace*{-0.1cm}
\epsfxsize = 3.0 in
\centerline{\epsfbox{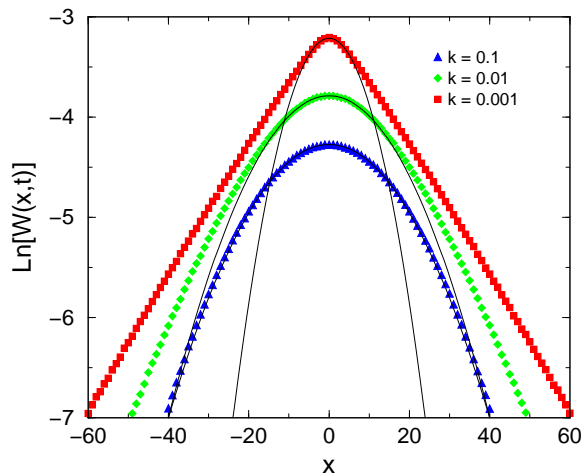}}
\protect\vspace*{0.2cm}
\caption{Dependence on $k$ of the size of the
Gaussian region at the center of the distribution.
As $k$ increases the width of the Gaussian region is increased. 
In the limit of $k=1$ the distribution becomes completely Gaussian. 
The data points represent the simulation and the solid lines represent 
the quadratic fits. Each simulation consisted of $ 10^7$ random walks, 
with $\epsilon = 1$, $D_0 = 1$ and $t = 256$.  
For clarity the distributions were shifted vertically and 
the tails are excluded from the plot.}
\label{figure5}
\end{figure}

\begin{figure}[htbp]
\protect\vspace*{-0.1cm}
\epsfxsize = 3.0 in
\centerline{\epsfbox{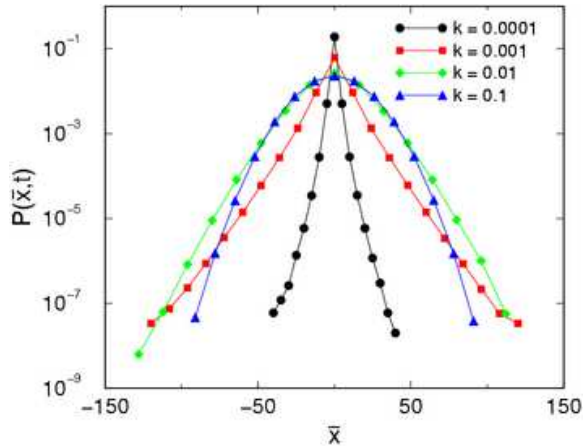}}
\protect\vspace*{0.2cm}
\caption{Distribution of $\bar{x}$ using the piecewise linear diffusion coefficient.  Note that the distribution depends on the parameter $k$. 
For small $k$, $\bar{x}$ becomes more localized around the origin.  
In this simulations, $\epsilon = 1$, $D_0 = 1$ and $t = 256$ with $10^7$ walks each. The solid lines are guides to the eye.}
\label{figure6}
\end{figure}

To understand the effects of a dynamic consensus value $\bar{x}$ on the distribution
of price returns, first consider the
case of piecewise linear diffusion $D(x-\bar{x},t)$. 
As discussed earlier, if $\epsilon = \frac{1}{\sqrt{D_0}}$ and $\bar{x}$ is static,
this form of the diffusion coefficient will result in an exponential return distribution.
Figs.~\ref{figure4} and~\ref{figure6} present the results of simulations with
dynamic $\bar{x}$. As expected from the argument in the previous paragraph, the
effect of the dynamics of $\bar{x}$ is to smooth out the peak in $W(x,t)$. In fact,
it becomes Gaussian in the center, as can be seen from the fit to the quadratic function
shown in red in Fig.~\ref{figure4}. That function is fit through the 31 points at 
the peak of $W(x,t)$.
The range of the quadratic region is directly related to the value of $k$.  If $k$ increases
this region is extended to a larger range, see Fig.~\ref{figure5}.
Away from the center of the distribution, where the effects of the dynamics of
$\bar{x}$ become less important, the exponential form of $W(x,t)$ is retained as expected.
This is shown by the fit to the line shown in blue, which works in the tail of the distribution.

Fig.~\ref{figure6} shows the distribution of $\bar{x}$ 
for different values of $k$, which we will call $P(\bar{x},t)$ 
to distinguish it from $W(x,t)$.
As $k$ is decreased, $x_0$ stays closer to the origin, its starting position, 
and the tails of the distribution decay rapidly.  This is why the tails of the
distribution of $x$ are not affected by the movement $\bar{x}$. In the limit of $k = 0$, $\bar{x}$
becomes static, and the distribution will be a single point at the origin.  
On the other hand, in the limit $k$ goes to 1 the distribution becomes a Gaussian.

\subsection{Quadratic Diffusion}

\begin{figure}[htbp]
\protect\vspace*{-0.1cm}
\epsfxsize = 3.0 in
\centerline{\epsfbox{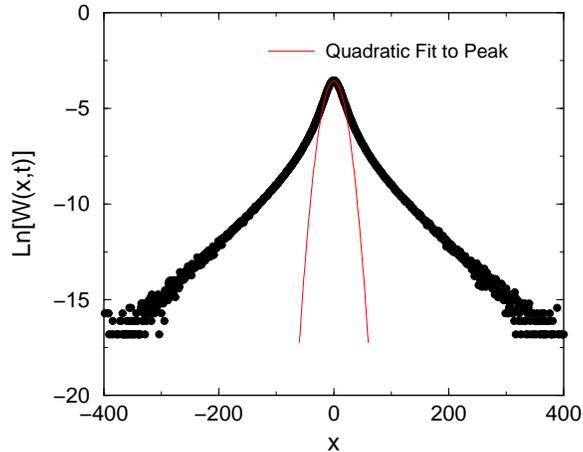}}
\protect\vspace*{0.2cm}
\caption{Distribution of $x$ using a dynamic consensus value and 
the quadratic diffusion coefficient. 
The central region of can be fitted using a quadratic function 
(solid line).  
To obtain the fitted line, 31 points at the center were used to fit 
the distribution.  
The data points show the results from the simulation using $\epsilon = 1$, $D_0 = 1$, $t = 256$, $k = 0.0005$ and $2 \times 10^7$ walks.}
\label{figure7}
\end{figure}

Now consider the effects of the dynamics of $\bar{x}$ on the return distribution
for the case where
$D(x-\bar{x},t)$ is a quadratic function.
In this case, as we have seen, if $\bar{x}$ is static, then the center of the re
turn distribution 
has a peak, but does not have discontinuity in the slope at $x=0$.  
Fig.~\ref{figure7} shows the return distribution calculated from simulations with dynamic $\bar{x}$.
As expected, the peak at $x=0$ is broader than in the case of static $\bar{x}$, 
and it can also be fitted with a quadratic function, indicating a Gaussian peak.
Notice, though, the tail behavior in this case differs from that observed 
when $\bar{x}$ was static. In this case, the tails of the distribution
are exponential.
We will return to this point at the end of this section.

\begin{figure}[htbp]
\protect\vspace*{-0.1cm}
\epsfxsize = 3.0 in
\centerline{\epsfbox{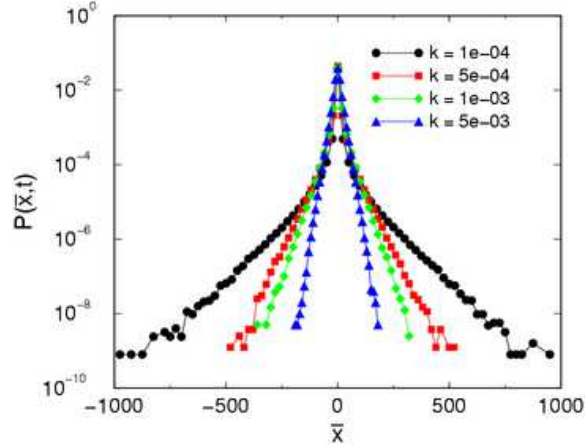}}
\protect\vspace*{0.2cm}
\caption{Distribution of $\bar{x}$ using the quadratic diffusion coefficient with dynamic consensus value.  The tails of the distribution get heavier as $k$ is decreased. This results were obtained using $\epsilon = 1$, $D_0 = 1$ and $t = 256$. The number of walks used was $2 \times 10^7$ for $k = 5 \times 10^{-3}$ and $k = 10^{-3}$, $4 \times 10^7$ for $k = 5 \times 10^{-4}$ and $5 \times 10^7$ for $k = 10^{-4}$. Here again the solid lines serve as guides to the eye.}
\label{figure8}
\end{figure}

Fig.~\ref{figure8} shows the distribution of $\bar{x}$.
As the value of $k$ decreases, the distribution of $\bar{x}$ becomes
sharply peaked.  This occurs at the same time that the tails of the
distribution are getting heavier, implying that compared to
Fig.~\ref{figure6} there is a larger chance of $\bar{x}$ being far
from its original position.  This behavior is presumably due to the effect
of the fat tails in the distribution of $x(t)$ for static $\bar{x}$.
When the value of $x(t)$ is in the tail of its distribution,
$\bar{x}$ is ``pulled'' far away from the origin.
This dynamics is very different than what we observed in the case of linear diffusion,
where the tails of the distribution of $\bar{x}$ decayed faster as $k$ decreased.
Notice that in the limit of $k = 0$ the distribution of $\bar{x}$ will also be a
single point at the origin, as is also the case for linear diffusion.

\begin{figure}[htbp]
\protect\vspace*{-0.1cm}
\epsfxsize = 3.0 in
\centerline{\epsfbox{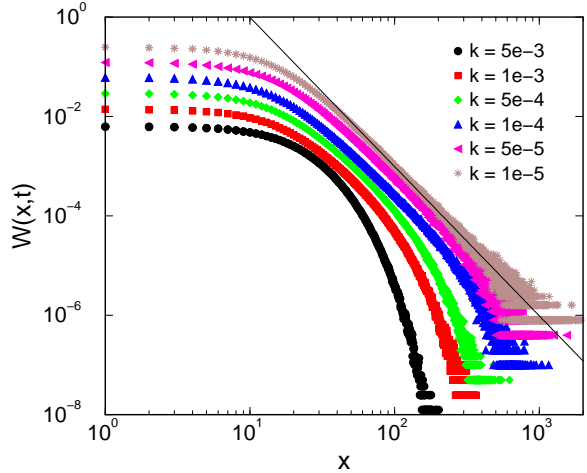}}
\protect\vspace*{0.2cm}
\caption{ Log-log plot of the distribution of $x$ with quadratic diffusion coefficient and dynamic consensus value.  The fat tails in the distribution go to a power law as $k$ goes to zero.
Each simulation used $\epsilon = 1$, $D_0 = 1$, $t = 256$ and $2 \times 10^7$ random walks. The black line at the top has a slope = 3. The distributions have been shifted vertically for clarity.}
\label{figure9}
\end{figure}

We now take a closer look at the tails of $W(x,t)$.  
Fig.~\ref{figure9} presents a log-log plot with results 
from simulations for quadratic diffusion using different
values of $k$ and $\epsilon = 1$.  
It is observed that as $k$ decreases, the power-law behavior starts to 
emerge in the tails.  In the limit of $k=0$ the results should be the 
same as in the static $\bar{x}$ case (a power-law with slope 3).  
To explain why the power-law disappear with an increase of the 
parameter $k$ we turn to the dynamics of $\bar{x}(t)$.
When $k$ is increased the value of $\bar{x}$ will follow
closer $x(t)$ making $D(x-\bar{x},t)$ have a more constant
value.  This results in the tails of $x(t)$ becoming Gaussian.

\section{Conclusions}

We have presented a theory for the distribution of stock returns.  
It is based on the conjecture
that the rate of trading of a stock depends on how far its current price is 
from a consensus price, $\bar{S}$.  
The resulting models use a non-constant diffusion coefficient $D(x,t)$
to simulate the rate
of returns.  
When $\bar{S}$ is fixed and a piecewise linear coefficient is used,
an exponential distribution of returns is found.
With quadratic
diffusion distributions with fat tails are found. 
The exponents describing
the power law fat tail distributions range from 2 to $\infty$. 
In both cases we obtained
an exact solution for $W(x,t)$ and simulations that support our findings.  
When $\bar{S}$ is allowed
to move, both forms of diffusion coefficient give 
distributions with an approximately 
Gaussian near the origin. 
Finally, we note that the range of behaviors observed
here with this simple model covers the range of non-Gaussian
behaviors seen in the distribution of returns of real financial markets.

A.L.A. and G.H.G are partially supported by the Institute for Space Science
Operations (ISSO) at the University of Houston. 
K.E.B. is supported by the NSF through grants DMR-0406323 and DMR-0427538,
and by the Alfred P. Sloan Foundation. 
G.H.G. is supported by the NSF through grant PHY-0202001.


\begin{thebibliography}{99}
\bibitem{levi} Possibilities that include anomalous diffusion have been considered 
in, for example, Refs.~\cite{manAsta} and \cite{conApot}.
\end{thebibliography}
\end{document}